\documentclass[a4paper,USenglish,autoref]{lipics-v2019}
\hideLIPIcs
\nolinenumbers

\usepackage{microtype}
\usepackage{amsmath}
\usepackage{amsthm}
\usepackage{amssymb}
\usepackage{xspace}

\usepackage{color}

\RequirePackage[normalem]{ulem} 
\RequirePackage{color}\definecolor{RED}{rgb}{1,0,0}\definecolor{BLUE}{rgb}{0,0,1} 

\newcommand{\BeginMyItemize}{\begin{itemize}}
\newcommand{\EndMyItemize}{\end{itemize}}

\newcommand{\BeginMyEnumerate}{\begin{enumerate}\setlength{\itemsep}{-\parskip}}
\newcommand{\EndMyEnumerate}{\end{enumerate}}

\renewcommand{\leq}{\leqslant}
\renewcommand{\geq}{\geqslant}

\newcommand{\Reals}{\mathbb{R}}

\newcommand{\dist}{\mathrm{dist}}

\renewcommand{\preceq}{\preccurlyeq}

\newcommand{\Exp}{\mathrm{Exp}}
\DeclareMathOperator{\degr}{degree}

\newcommand{\PP}{\ensuremath{\mathbb{P}}}
\newcommand{\EE}{\ensuremath{\mathbb{E}}}

\newcommand{\ceil}[1]{\left\lceil #1 \right\rceil}
\newcommand{\ceili}[1]{\lceil #1 \rceil}
\newcommand{\br}[1]{\left( #1 \right)}
\newcommand{\sbr}[1]{\left[ #1 \right]}
\newcommand{\floor}[1]{\left\lfloor #1 \right\rfloor}
\newcommand{\floori}[1]{\lfloor #1 \rfloor}

\newcommand{\mydef}{:=}
\newcommand{\etal}{\emph{et al.}}

\newcommand{\tspp}{\mbox{{\sc Traveling Salesman Problem}}\xspace}

\newcommand{\mrst}{\mbox{{\sc Minimum Rectilinear Steiner Tree}}\xspace}

\newcommand{\walls}{\mathcal{W}}
\newcommand{\myhard}{\mathrm{hard}}  
\newcommand{\hwalls}{\walls_{\myhard}}
\newcommand{\mysoft}{\mathrm{soft}}  
\newcommand{\swalls}{\walls_{\mysoft}}

\newcommand{\labda}{\lambda}

\newcommand{\sqrtd}{\sqrt{\delta}}

\newcommand{\csqrtd}{\ceil{\sqrtd}}
\newcommand{\cisqrtd}{\ceili{\sqrtd}}

\newcommand{\mypara}[1]{\medskip\noindent{\sf\textbf{#1}}}

\definecolor{darkgreen}{rgb}{0.0, 0.5, 0.0}

\usepackage{algorithm}
\usepackage[noend]{algpseudocode}

\title{Rectilinear Steiner Trees in Narrow Strips}
\funding{The work in this paper is supported by the Netherlands Organisation for Scientific Research (NWO) through Gravitation-grant NETWORKS-024.002.003.}
\author{Henk Alkema}{Department of Mathematics and Computer Science, TU Eindhoven, the Netherlands}{h.y.alkema@tue.nl}{}{}
\author{Mark de Berg}{Department of Mathematics and Computer Science, TU Eindhoven, the Netherlands}{m.t.d.berg@tue.nl}{}{}

\authorrunning{H.~Alkema and M.~de Berg} 

\Copyright{Henk Alkema and Mark de Berg}



\EventEditors{John Q. Open and Joan R. Access}
\EventNoEds{2}
\EventLongTitle{42nd Conference on Very Important Topics (CVIT 2016)}
\EventShortTitle{CVIT 2016}
\EventAcronym{CVIT}
\EventYear{2016}
\EventDate{December 24--27, 2016}
\EventLocation{Little Whinging, United Kingdom}
\EventLogo{}
\SeriesVolume{42}
\ArticleNo{23}

\theoremstyle{plain}
\newtheorem{observation}[theorem]{Observation}

\begin{document}
\maketitle

\begin{abstract}
A \emph{rectilinear Steiner tree} for a set $P$ of points in $\Reals^2$ is a tree that connects the points in $P$ using horizontal and vertical line segments.
The goal of \mrst is to find a rectilinear Steiner tree with minimal total length.
We investigate how the complexity of \mrst for point sets $P$ inside the strip $(-\infty,+\infty)\times [0,\delta]$ depends on the strip width~$\delta$.
We obtain two main results.
\begin{itemize}
\item We present an algorithm with running time $n^{O(\sqrtd)}$ for sparse point sets, 
      that is, point sets where each $1\times\delta$ rectangle inside the strip contains $O(1)$ points.
\item For random point sets, where the points are chosen randomly inside a rectangle of height~$\delta$
      and expected width~$n$, we present an algorithm that is fixed-parameter tractable with respect 
      to~$\delta$ and linear in $n$. It has an expected running time of $2^{O(\delta \sqrtd)} n$.
\end{itemize}
\end{abstract}

\section{Introduction}
In the {\sc Minimum Steiner Tree} problem in the plane, we are given as 
input a set $P$ of points in the plane, called \emph{terminals}, and the goal 
is to find a minimum-length tree that connects the terminals in $P$.
Thus the given terminals must be nodes of the tree, but the tree may
also use so-called \emph{Steiner points} as nodes.
{\sc Minimum Steiner Tree} is a classic optimization problem. It was
among the first problems to be proven NP-hard, not only for the case where
the length of the tree is measured using Euclidean metric~\cite{10.2307/2100193}
but also in the rectilinear version~\cite{10.2307/2100192}.
It was also shown to be NP-hard for other metrics~\cite{DBLP:journals/ijcga/BrazilZ14}

The rectilinear version of the problem, where the edges of the tree must be 
horizontal or vertical, is one of the most widely studied variants, and it is also the topic of our paper. 
The \mrst problem dates back more than 50~years~\cite{10.2307/2099400, 10.2307/2946265}. 
Its popularity arises from its many applications, in particular in the design of integrated 
circuits~\cite{BrazilZachariasen15, DBLP:journals/algorithmica/BrazilTWZ06, DBLP:journals/jgo/BrazilZ09, DBLP:journals/siamcomp/WidmayerWW87}.
The two most important early insights on \mrst came from Hanan~\cite{10.2307/2946265} 
and  Hwang~\cite{10.2307/2100587}. Hanan observed that any terminal set $P$ admits a
minimum rectilinear Steiner tree (MRST, for short) whose edges lie on the grid formed 
by all horizontal and vertical lines passing through at least one terminal in $P$. 
This grid is often called the \emph{Hanan grid}. This implies that the
\mrst problem can be reduced to a purely combinatorial problem---namely, a Steiner-tree
problem on graphs---which is not possible for the Euclidean version of the problem.
Hwang investigated the structure of optimal MRSTs in more detail, 
by providing a characterization of the different components of an MRST; see Section~\ref{sec:preli}.

As mentioned, \mrst can be considered a special case of the Steiner-tree problem on graphs.
Here the input is an edge-weighted graph $G=(V(G),E(G))$ and a terminal set $P\subseteq V(G)$, 
and the goal is to compute a minimum-length subtree of $G$ that includes all terminals.
In~1971 Dreyfus and Wagner~\cite{DBLP:journals/networks/DreyfusW71} gave an algorithm 
solving the Steiner-Tree problem on graphs in  time $3^n \cdot \log W \cdot |V(G)|^{O(1)}$, 
where $W$ is the maximum edge weight in $G$. This was later improved by 
Bj\"orklund~\etal~\cite{DBLP:conf/stoc/BjorklundHKK07} and Nederlof~\cite{DBLP:journals/algorithmica/Nederlof13},
who gave an algorithm with $2^n \cdot W \cdot |V(G)|^{O(1)} $ running time.
A variant of the Dreyfus-Wagner algorithm for \mrst runs in time $O(n^2 \cdot 3^n)$.
Thobmorson~\etal~\cite{DBLP:conf/paa/ThomborsonDS87} and Deneen~\etal~\cite{randomMRST} 
gave randomized algorithms for the special case of \mrst where the terminals are drawn independently 
and uniformly from a rectangle. Both run in $2^{O(\sqrt{n} \log n)}$ expected time.
Finally, in 2018 Fomin~\etal~\cite{rectiSteiner16} presented a $2^{O(\sqrt{n} \log n)}$ algorithm for general point sets.

Due to the many applications of {\sc Minimum Steiner Tree} variants in the plane, there has also
been significant interest in practical implementations. These implementations rely on the
insight that a minimum Steiner tree can always be decomposed into so-called \emph{full components}, 
which are maximal subtrees that do not have any terminals as internal nodes~\cite{10.2307/2100587}.
(This holds for the Euclidean as well as the rectilinear version.)
To compute an exact solution, a set of candidate full components is first computed and then
it is computed which subset of candidate full components can be concatenated into an MRST.
This process was introduced by Winter in 1985 \cite{DBLP:journals/networks/Winter85}, 
in his software package \emph{GeoSteiner}. Still, only very small data sets could be handled,
and even in 1994 the state-of-the-art software could solve the rectilinear variant of
the problem for only up to 16~points~\cite{DBLP:conf/dimacs/ThomborsonAC92}.
Warme's dissertation~\cite{warme-thesis} significantly improved the process of concatenating the full components, 
resulting in optimal Steiner trees for up to 1,000 points for the rectilinear version of the
problem and up to 2,000 points for the Euclidean version. In 1998 Althaus~\cite{Althaus}
obtained similar results.
Throughout the years, GeoSteiner, which had become a collaboration between Warme, Winter and Zachariasen, has remained 
the fastest publicly available software package for computing minimum Steiner trees in the plane.
By 2018, it could solve instances for up to 4,000 points for the rectilinear version, and
up to 10,000 points for the Euclidean version~\cite{geosteiner16}.

\mypara{Our contribution.}
The fastest known algorithm for \mrst in $\Reals^2$ runs 
in $2^{O(\sqrt{n} \log n)}$ time~\cite{rectiSteiner16}. In $\Reals$, on the other hand, 
the problem can be trivially solved in $O(n \log n)$ time by just sorting the
points. In order to better understand the computational complexity of the classic
\mrst problem in the plane, we therefore investigate how the complexity depends
on the width of the terminal set~$P$. If the point set in $P$ is ``almost 1-dimensional''
in the sense that the points lie in a narrow strip $\Reals \times [0,\delta]$, then can we solve \mrst
more efficiently than in the general case? And if so, how does the complexity scale
with~$\delta$? Can we obtain an algorithm that is fixed-parameter tractable
with respect to~$\delta$? This follows the line of research started recently by
Alkema~\etal~\cite{DBLP:conf/compgeom/AlkemaBK20}, who studied these questions for the \tspp.
We study these questions in the following two scenarios.
\begin{itemize}
\item \emph{Sparse point sets.}
    In this scenario, for any $x \in \Reals$ the rectangle $[x,x+1]\times [0,\delta]$ contains $O(1)$ points.
    We show that for sparse point sets in $\Reals^2$ an MRST must be $k$-tonic---an MRST is $k$-tonic 
    if it intersects any vertical line 
    at most $k$-times---for $k=O(\sqrtd)$, and we 
    give a dynamic-programming algorithm which runs in $n^{O(\sqrtd)}$ time.
\item \emph{Random point sets.}
    Our main result is for point sets $P$ generated randomly inside a rectangle of height~$\delta$
    and expected width~$n$, as follows. First, we generate $n$ independent exponentially distributed variables 
    $\Delta_0,...,\Delta_{n-1} \sim \Exp (1)$. Using these, we compute the $x$-coordinates of our points
    by setting $x_i$, the $x$-coordinate of the $i$-th point from $P$, as $x_i := \sum_{j=0}^{i-1} \Delta_i$ for $1 \leq i \leq n$. 
    Next, we generate the $y$-coordinates of the points by picking each $y_i$ uniformly 
    and independently from the interval~$[0,\delta]$. Thus the points from $P$ lie inside the rectangle $[0,x_n]\times[0,\delta]$.
    One can show that asymptotically this distribution is essentially the same as the distribution obtained
    by picking $n$ points uniformly at random from the rectangle~$[0,n]\times [0,\delta]$~\cite{daley2007introduction}.
    However, the  random point process as just described is somewhat easier to analyze,
    so we will assume the points are generated according to that process.
    For this case we provide an FPT algorithm for \mrst, which runs in expected time $2^{O(\delta \sqrtd)}n$.
    More precisely, expected running time is $\min (n^{O(\sqrt{\delta})}, 2^{O(\delta \sqrtd)}n)$.
    Note that the running time is linear when $\delta=O(1)$. 
\end{itemize}

\section{Preliminaries} 
\label{sec:preli}

\mypara{Notation and terminology.}
Let $P:=\{p_1,\ldots,p_n\}$ be a set of \emph{terminals} in a $2$-dimensional strip with height $\delta$ ---we call such a strip a \emph{$\delta$-strip}---which we assume without loss of generality to be~$\Reals\times [0,\delta]$.
We use $x_i$ and $y_i$ to denote the $x$- and $y$-coordinate of point $p_i$, respectively.
The points can be easily sorted on their $x$-coordinates: this can be done in $O(n \log n)$ time for sparse point sets, and in $O(n)$ expected time for random point sets \cite{clrs-ita-09}.
Therefore, we will from now on assume that $x_i \leq x_j$ for all $1\leq i \leq j \leq n$.
We define the \emph{spacing} of $p_i$ (in $P$) as $\Delta_i \mydef x_{i+1} - x_i$, for all $1 \leq i \leq n-1$.
We write $P[i,j]$ to denote the set $\{p_i,...,p_j\}$. We denote the vertical
distance between two horizontal edges $e,e'$ (or the horizontal distance between two
vertical edges) by~$\dist(e,e')$.
\medskip

\begin{figure}
\begin{center}
\includegraphics{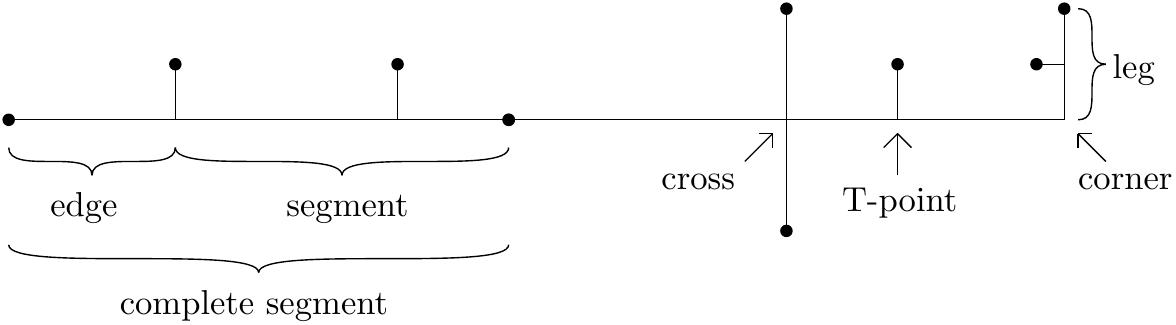}
\caption{Illustration of terminology concerning rectilinear Steiner trees}
\label{fig:prelims:terminology1}
\end{center}
\end{figure}

Next we give some (mostly standard) terminology concerning rectilinear Steiner trees;
see also Figure~\ref{fig:prelims:terminology1}.
A \emph{rectilinear tree} is a tree structure embedded in the plane whose edges are horizontal or vertical line segments overlapping only at their endpoints.
The \emph{length} of a tree $T$, or $\|T\|$, is the sum of the lengths of its edges.
A \emph{rectilinear Steiner tree} for a set $P$ of terminals is a rectilinear tree such that each terminal $p \in P$ is an endpoint of an edge in the tree.
A \emph{minimal rectilinear Steiner minimal tree (MRST)} is such a tree of minimum length.

The \emph{degree} of a (Steiner or terminal) point~$q$ in a tree $T$ is the number of edges incident on it.
We denote the degree of $q$ in $T$ by $\degr_T(q)$, or simply $\degr(q)$ when $T$ is clear from the context.
Without loss of generality, if a degree-2 point has collinear (i.e. both horizontal or both vertical) incident edges then that point must be a terminal.
Clearly, a point has degree at most $4$.
A point with degree of at least $3$ that is not a terminal is called a \emph{Steiner point}.
A \emph{corner} is a degree-$2$ point with non-collinear incident edges that is not a terminal.
Hence, each endpoint of an edge is either a terminal, a Steiner point, or a corner.

A \emph{segment} is defined to be a sequence of one or more adjacent collinear edges, with no terminals in the 
segments' interior.\footnote{When we refer to the ``interior'' of a segment, we always mean its relative interior, i.e. the segment excluding its endpoints.}
A \emph{complete segment} is a inclusion-wise maximal segment.
Note that a complete segment does not have terminals in its interior.
A corner is incident to exactly one horizontal complete segment and exactly one vertical complete segment.
These complete segments are the \emph{legs} of the corner.
A \emph{T-point} is a degree-3 Steiner point.
Finally, a \emph{cross} is a degree-4 Steiner point.
Note that the endpoints of a complete segment are T-points, corners or terminals.
\medskip

Separators will play a crucial role in our algorithms.
A \emph{separator} is a vertical line, not containing any of the points in $P$, that separates~$P$ into two non-empty subsets.
For all $1 \leq i < n$ such that $x_i < x_{i+1}$, we define $s_i$ to be the separator with $x$-coordinate $(x_i + x_{i+1})/2$.
The \emph{tonicity} of a rectilinear tree $T$ \emph{at a separator} $s$ is the number of times $T$ crosses $s$;
when the tonicity of $T$ at $s$ is~1, we call it \emph{monotonic} at~$s$.
The \emph{tonicity} of a rectilinear tree $T$ is the maximum over the tonicity of $T$ at all separators.
A rectilinear tree is called \emph{monotonic} when its tonicity is~1.

\mypara{Characterisation of the MRST.}
Over the years, many different properties of the MRST have been proven.
One of the most important ones is the following:
\begin{observation}[Hanan~\cite{10.2307/2946265}]\label{obs:prelim:hanan}
Let $P$ be a set of terminals in $\Reals^2$.
Then there exists an MRST on $P$ that is a subset of the \emph{Hanan grid}, the grid formed by taking all horizontal and vertical lines which pass through at least one of the points of $P$.
\end{observation}
From now on, we will only consider rectilinear Steiner trees that lie on the Hanan grid.
Furthermore, we can now directly conclude that the tonicity of an MRST is at most $n$.

\begin{figure}
\begin{center}
\includegraphics{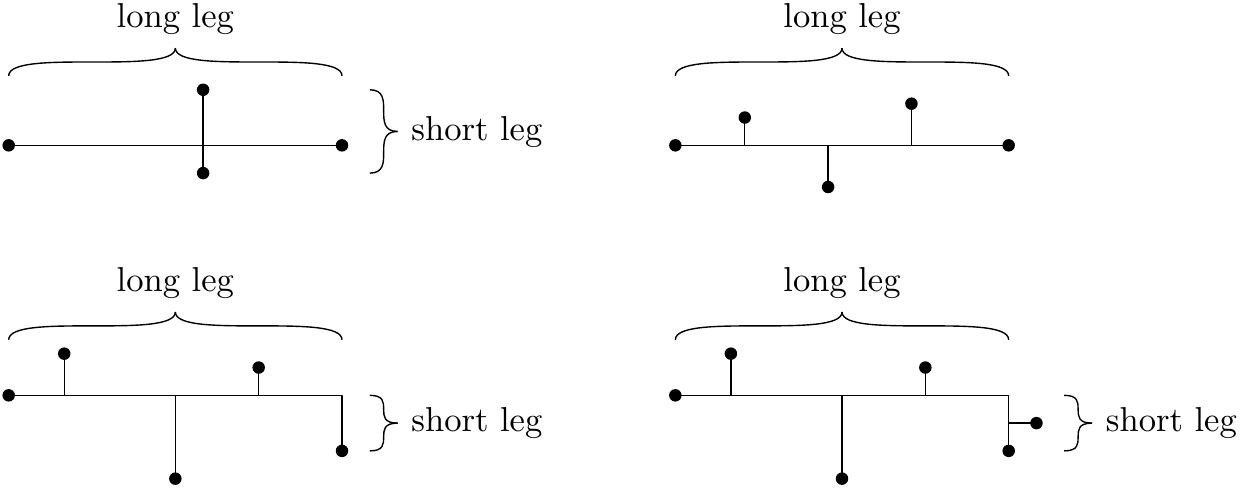}
\caption{An example of each of the four different types of full components in Hwang trees.}
\label{fig:prelims:hwangtypes}
\end{center}
\end{figure}

\begin{figure}
\begin{center}
\includegraphics{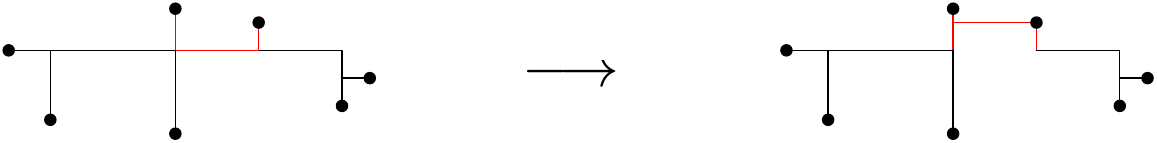}
\caption{An example showing that Hwang trees do not contain a U-shape.
The tree on the left has the same length as the tree on the right, but contains fewer full components.
Therefore, the tree on the left is not a Hwang tree.}
\label{fig:prelims:hwangU}
\end{center}
\end{figure}

A continuation on this characterisation is given by the Hwang theorem.
We define a \emph{full component} of a rectilinear Steiner tree $T$ to be a maximal subtree that does not have any terminals as internal nodes.
Note that a node in a full component of an MRST is a terminal if and only if it is a leaf in that component.
Also note that any terminal $p_i\in P$ will be a leaf in exactly $\mbox{degree}(p_i)$ full components.
Hwang's theorem is now given by the following:
\begin{theorem}[Hwang~\cite{10.2307/2100587}]
Let $P$ be a set of terminals in $\Reals^2$.
Then there exists an MRST $T$ on $P$ with a maximal number of full components, such that each full component $C$ is of one of the following four types.
Let $m_C$ be the number of terminals in $C$.  
Then $C$ consists of
\begin{itemize}
    \item four edges, connected in a cross,
    \item a single complete segment with $m_C-2$ alternating incident edges,
    \item a corner and its legs, with $m_C-2$ alternating edges incident to a single leg, or
    \item a corner and its legs, with $m_C-3$ alternating edges incident to a single leg and a single edge incident to the other leg.
\end{itemize}
For all legs, the incident edge closest to the corner must point away from the opposite leg.
Furthermore, the edges incident to the long leg on the same side as the short leg are at least as long as the short leg.
\end{theorem}
We will call MRSTs which have this property \emph{Hwang trees}.
See Figure~\ref{fig:prelims:hwangtypes} for an example of each of the four types of full components of Hwang trees.
Note that these full components do not contain a U-shape formed by an edge and two adjacent segments lying to the same side of that edge;
any component with such a  U-shape can be split into two full components by sliding the edge towards the terminals at the end of those segments.
See Figure~\ref{fig:prelims:hwangU} for an example.
We will call the complete segment with the $m_C-2$ or $m_C-3$ incident edges the \emph{long leg}, and the other leg (if any) the \emph{short leg}.
If there are two complete segments which both have $m_C-2$ or $m_C-3$ incident edges, we will consider the horizontal one to be the long leg, and the vertical one to be the short leg.
If the long leg is horizontal (vertical), we call the full component a \emph{horizontal (vertical) full component}.


\section{Sparse point sets inside a narrow strip} \label{sec:sparse}
We say a point set is \emph{sparse} if for all $x$ the rectangle $[x,x+1] \times [0,\delta]$ 
contains at most $k$ points for some arbitrary but fixed \emph{sparseness constant}~$k$.
In this section, we will give a $n^{O(\sqrtd)}$ algorithm for sparse point sets.
We will do so in two steps.
First, we will show that all separators are crossed at most $O(\sqrtd)$ times.
Then, we will give a dynamic-programming algorithm which sweeps from left to right and runs in the desired time.

First, we will show that parallel edges of an MRST cannot be too close.
Recall that $\Delta_i$ denotes the horizontal spacing between $p_i$ and $p_{i+1}$,
and that $\delta$ denotes the height of the strip containing~$P$.
Also recall that $s_i$ is the separator in between the points $p_i$ and $p_{i+1}$.
\begin{observation}\label{obs:gen:paralleledges}
(i) Let $E = \{e_1,\ldots,e_m\}$ be a set of $m$ horizontal edges of an MRST~$T$ which 
all intersect two vertical lines $\ell$ and $\ell'$.
Then $m \leq 1+\floori{\delta/ \dist(\ell,\ell')}$.
A similar statement holds when $E$ is a set of vertical edges intersecting two horizontal lines.\\
(ii) If $\Delta_i > \delta$, then the tonicity of any MRST at $s_i$ is~1.
\end{observation}
\begin{proof}
We will first prove (i).
W.l.o.g., let the edges in $E$ be numbered from top to bottom, and let $\ell$ lie to the left of $\ell'$.
Suppose for a contradiction that $m > 1 + \floori{\delta/\dist(\ell,\ell')}$.
Since $\dist(e_1,e_m) \leq \delta$, there are two edges $e_{i}$ and $e_{i+1}$ such that $\dist(e_i,e_{i+1}) \leq \delta / (m-1) < \dist(\ell,\ell')$.
We will now create a rectilinear Steiner tree~$T'$ strictly shorter than $T$, giving the desired contradiction.
To this end we first delete the part of $e_{i}$ between $\ell$ and $\ell'$. 
Let $e_{i,1}$ denote the part of $e_i$ to the left of $\ell$ (if any) and let $e_{i,2}$ denote the part of $e_i$ to the right of $\ell'$ (if any);
see Figure~\ref{fig:gen:paralleledges}.
\begin{figure}
\begin{center}
\includegraphics{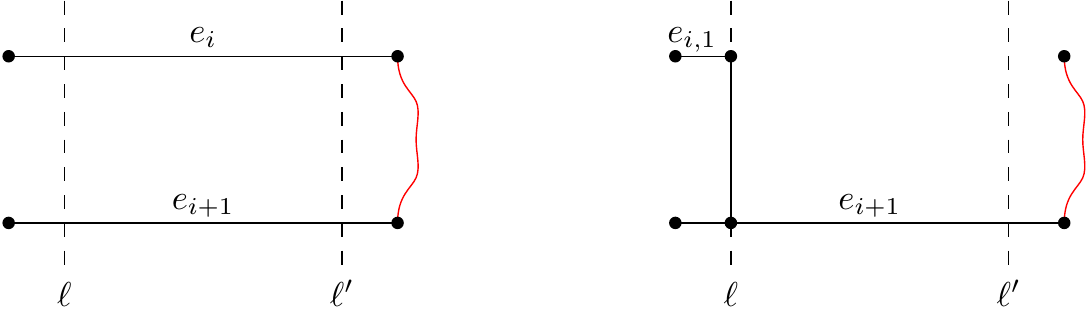}
\caption{Illustration for the proof of Observation~\ref{obs:gen:paralleledges}.
On the left, $T$. On the right, $T'$.
Since the deleted part of $e_i$ is longer than the vertical edge connecting
$e_{i,1}$ to $e_{i+1}$, the tree $T'$ is shorter than~$T$.}
\label{fig:gen:paralleledges}
\end{center}
\end{figure}
The deletion splits $T$ into two components. Assume without loss of generality that
$e_{i,2}$ is in the same component as $e_{i+1}$. By deleting $e_{i,2}$ and connecting
$e_{i,1}$ to $e_{i+1}$ with a vertical edge contained in~$\ell$, we create a rectilinear Steiner Tree $T'$ such that
\[
\|T'\| = \|T\| - \dist(\ell,\ell') - |e_{i,2}| + \dist(e_i, e_{i+1}) < \|T\|,
\]
giving the desired contradiction.

Part~(ii) of the observation directly follows from part~(i).
To see this, let $\Delta_i > \delta$ for some $i$.
Then there are two lines $\ell$ and $\ell'$ between $p_i$ and $p_{i+1}$ such that $\dist(\ell,\ell') > \delta$.
Let $T$ be an MRST.
Note that any edge of $T$ that crosses $s_i$ also crosses $\ell$ and $\ell'$.
Therefore, by part~(i) we know that $T$ crosses $s_i$ at most $m \leq 1 + \floori{\delta/\dist(\ell,\ell')} = 1$ times.
\end{proof}

We are now ready to bound the tonicity at the separators.
The following lemma will also be applicable for randomly generated point sets.
\begin{lemma}\label{lem:gen:sqrtdeltacrossings}
Let $T$ be a Hwang tree on $P$. 
Let $s_i$ be a separator such that 
\[
x_{i+\cisqrtd+c_1} - x_{i} > c_2 \sqrtd
\]
for an integer constant $c_1 \geq 0$ and a constant $c_2 > 0$.
Then the tonicity of $T$ at $s_i$ is~$O(\sqrtd+1)$.
\end{lemma}
\begin{proof}
We will show that $T$ crosses $s_i$ at most $17 c_1 + 36 + (4/c_2 + 17) \sqrtd$ times. 
Recall that $P[i,j] \mydef \{p_i,...,p_j\}$ and define $P' \mydef P[i,i+\cisqrtd+c_1]$.
The edges of $T$ crossing $s_i$ can be split into five sets:
\begin{itemize}
    \item $E_{\mathrm{long}}$, the set of edges which also cross the vertical line defined by $x = x_i + c_2 \sqrtd / 2$.
    \item $E_{\mathrm{h}}$, the set of edges not in $E_{\mathrm{long}}$ that are part of a horizontal full component.
    \item $E_{\mathrm{vl}}$, the set of edges not in $E_{\mathrm{long}}$ that are part of a vertical full component whose long leg lies to the left of $s_i$.
    \item $E_{\mathrm{vb}}$, the set of edges not in $E_{\mathrm{long}}$ that are part of a vertical full component whose long leg lies between $s_i$ and the vertical line $x = x_i + c_2 \sqrtd$.
    \item $E_{\mathrm{vr}}$, the set of edges not in $E_{\mathrm{long}}$ that are part of the short leg of a vertical full component whose long leg lies to the right of the vertical line $x = x_i + c_2 \sqrtd / 2$.
\end{itemize}
See Figure~\ref{fig:gen:sqrtdcrossings} for examples.
\begin{figure}
\begin{center}
\includegraphics{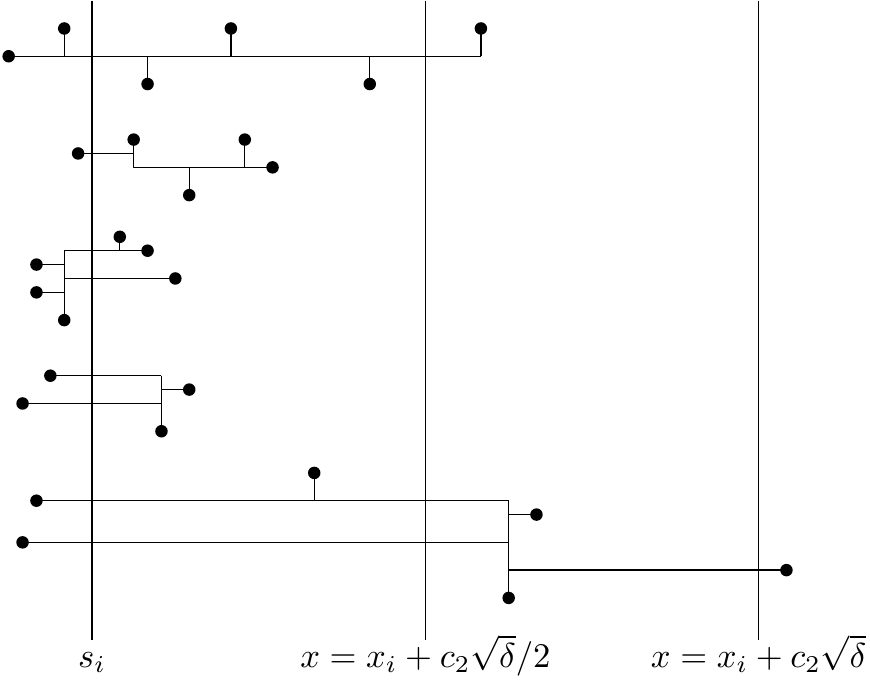}
\caption{Examples of the different types of edges crossing $s_i$, not to scale.
Topmost, two edges in $E_{\mathrm{h}}$, one as part of the long leg, and one as the incident edge of the short leg of a horizontal full component.
Then, two edges in $E_{\mathrm{vl}}$, one as part of the short leg, and one as an edge incident to the long leg.
Third, two edges in $E_{\mathrm{vb}}$, one as the short leg, and one as an edge incident to the long leg.
Finally, an edge in $E_{\mathrm{vr}}$ and an edge in $E_{\mathrm{long}}$.}
\label{fig:gen:sqrtdcrossings}
\end{center}
\end{figure}
By Observation~\ref{obs:gen:paralleledges}, we have
\[|E_{\mathrm{long}}| \leq 1+\floor{\delta / \br{c_2 \sqrtd / 2}} \leq  1+2 \sqrtd / c_2.\]
Secondly, every horizontal full component corresponding to a edge in $E_{\mathrm{h}}$ contains a terminal from $P'$.
Since horizontal full components cross $s_i$ at most once and all terminals are part of at most four full components, we conclude that
\[|E_{\mathrm{h}}| \leq 4 |P'|.\]
The right endpoints of edges in $E_{\mathrm{vl}}$ are either terminals in $P'$ or T-points incident to a vertical edge whose other endpoint is a terminal in $P'$.
Since all terminals are part of at most four full components, we conclude that
\[|E_{\mathrm{vl}}| \leq 4 |P'|.\]
To bound $|E_{\mathrm{vb}}|$, let $C_{\mathrm{vb}}$ be the set of all full components with at least one edge in~$E_{\mathrm{vb}}$.
For a component $C \in C_{\mathrm{vb}}$, let $S(C)$ be the set of horizontal complete segments to the right of the long leg of $C$.
Recall that the segments incident to the long leg of a  vertical full component $C$ alternate between lying to the right and left of the long leg.
Hence, the number of edges in $E_{\mathrm{vb}}$ from $C$ is bounded by $|S(C)| + 1$, and so
\[|E_{\mathrm{vb}}| \leq |C_{\mathrm{vb}}| + \sum_{C \in C_{\mathrm{vb}}} |S(C)|.\]
Note that for any component $C \in C_{\mathrm{vb}}$, the terminal incident to its long leg 
must be a point in $P'$. Furthermore, every point in $P'$ can only be used twice this way.
Therefore, $|C_{\mathrm{vb}}| \leq 2 |P'|$.

To bound  $\sum_{C \in C_{\mathrm{vb}}} |S(C)|$, we note that each segment in any set $S(C)$
either 
(i) is a short leg,
(ii) ends in a point of $P'$, or 
(iii) crosses the vertical line $x = x_i + c_2 \sqrtd$.
Since $|C_{\mathrm{vb}}| \leq 2 |P'|$, there are at most $2 |P'|$ segments of type (i).
Trivially, there are at most $|P'|$ segments of type (ii).
Finally, by Observation~\ref{obs:gen:paralleledges}, we can only have $1+\floori{\delta / (c_2 \sqrtd / 2)} \leq  1+2 \sqrtd / c_2$ segments of type (iii).
Therefore, 
\[\sum_{C \in C_{\mathrm{vb}}} |S(C)| \leq 3 |P'| + 1+2 \sqrtd / c_2.\]
We conclude that 
\[|E_{\mathrm{vb}}| \leq 2|P'| + 3 |P'| + 1+2 \sqrtd / c_2 = 5|P'| + 1+2 \sqrtd / c_2.\]
Finally, the right endpoints of edges in $E_{\mathrm{vr}}$ are T-points incident to a vertical edge whose other endpoint is a terminal in $P'$.
Since all terminals are part of at most four full components,
\[|E_{\mathrm{vr}}| \leq 4 |P'|.\]
Since $|P'|=\ceil{\sqrtd}+c_1+1$, the total number of edges crossing $s_i$ can now be bounded by
\begin{align*}
|E_{\mathrm{long}}| + |E_{\mathrm{h}}| + |E_{\mathrm{vl}}| + |E_{\mathrm{vb}}| + |E_{\mathrm{vr}}|
&\leq 17|P'| + 2(1+2 \sqrtd / c_2) \\
&\leq 17 c_1 + 36 + (4/c_2 + 17) \sqrtd. \qedhere
\end{align*} 
\end{proof}

Using Lemma~\ref{lem:gen:sqrtdeltacrossings} we can now prove a bound on the tonicity of MRSTs of sparse point sets.
\begin{corollary}\label{cor:sparse:sqrtdton}
An MRST on a sparse point set~$P$ in a $\delta$-strip is  $\br{(9k+18)(2+ \sqrtd)}$-tonic,
where $k$ is the sparseness constant.
\end{corollary}
\begin{proof}
First, we note that since our point set $P$ is sparse, we have $x_j - x_i \geq \floori{(j-i)/k}$ for all $j>i$.
Specifically, for all $s_i$ such that $i+\cisqrtd+k \leq n$, we get
\[x_{i+\csqrtd+k} - x_i \geq \floor{\frac{\csqrtd+k}{k}} \geq \floor{\frac{\sqrtd}{k}+1}\geq \frac{\sqrtd}{k}.\]
Therefore, we can invoke Lemma~$\ref{lem:gen:sqrtdeltacrossings}$ with $c_1 = k$ and $c_2 = 1/k$, 
giving us that all these $s_i$ are crossed at most $17k + 36 + (4k + 17) \sqrtd$ times.
(These constant follow from the constants in the proof of Lemma~$\ref{lem:gen:sqrtdeltacrossings}$,
see the Appendix.)
By symmetry, we can do the same for all $s_i$ such that $(i+1)-\cisqrtd-k \geq 1$.
Finally, we note that if this does not cover all $s_i$, then we have fewer than $17k + 36 + (4k + 17) \sqrtd$ points in total.
Since every separator is crossed at most $n$ times, the statement also holds in this case.
We conclude that for sparse terminal sets, all separators are crossed at most $17k + 36 + (4k + 17) \sqrtd < (9k+18)(2+ \sqrtd)$ times.
\end{proof}

\begin{figure}
\begin{center}
\includegraphics{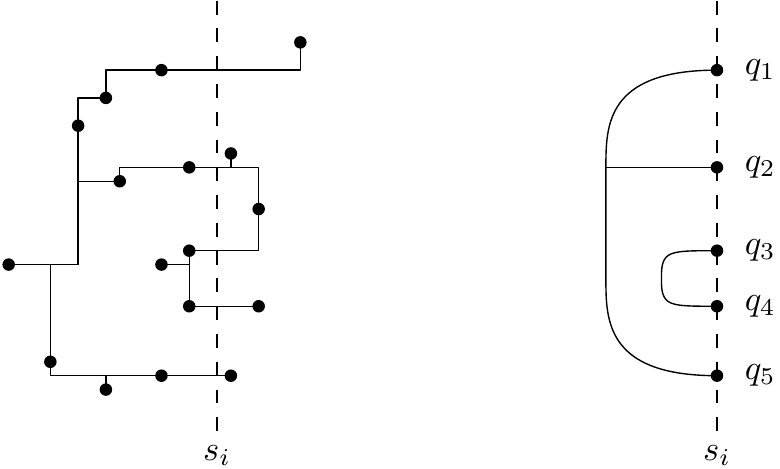}
\caption{An example of an MRST and its crossing pattern $C = \{ \{q_1,q_2,q_5\}, \{q_3,q_4\} \}$ at $s_i$ }
\label{fig:sparse:crossingpattern}
\end{center}
\end{figure}

Corollary~\ref{cor:sparse:sqrtdton} gives rise to a natural dynamic-programming algorithm, 
as explained next. Let $T$ be a rectilinear Steiner tree, and let $s_i$ be a separator.
We define the $\emph{crossing pattern}$ of $T$ at $s_i$ as follows.
Let $X(s_i)$ be the set of at most $n$ points where the Hanan grid crosses~$s_i$, and let $X(s_i,T) \subseteq X(s_i)$ be the subset of points where $T$ crosses~$s_i$.
If $T$ is an MRST,
\[
|X(s_i,T)|\leq (9k+18)(2+ \sqrtd)=O(\sqrtd)
\]
by Corollary~\ref{cor:sparse:sqrtdton}.
We partition $X(s_i,T)$ into parts (that is, subsets) such that two points from $X(s_i,T)$ 
are in the same part if the path in $T$ between these points fully lies to the left of~$s_i$.
The resulting partition of $X(s_i,T)$ is the crossing pattern of $T$ at~$s_i$;
see Figure~\ref{fig:sparse:crossingpattern} for an example.
We will say that a rectilinear forest $T$ \emph{adheres to} $C$ at $s_i$ if $T$ lies fully to the left of~$s_i$, 
and there exists a rectilinear forest $T'$ which lies fully to the right of $s_i$ such that $T \cup T'$ 
is a rectilinear Steiner tree with crossing pattern $C$ at~$s_i$.
Note that not all crossing patterns can lead to an MRST:
those that require crossing edges on the left-hand side (because
they do not have a proper `nesting structure') can never lead to an MRST.
We call the crossing patterns that contain at most $(9k+18)(2+ \sqrtd)$ points and do not require crossing edges on the left-hand side \emph{viable crossing patterns}.
We will now count the number of viable crossing patterns at $s_i$.
There are $n^{O(\sqrtd)}$ possible sets $X(s_i,T)$ that contain at most $(9k+18)(2+ \sqrtd)$ points.
The number of viable partitions of these points---also known as the number of non-crossing partitions---follows the Catalan numbers.
Hence, there are $2^{O(\sqrtd)}$ possible viable partitions for each $X(s_i,T)$.
This implies that the total number of viable crossing patterns for $s_i$ is $n^{O(\sqrtd)} \cdot 2^{O(\sqrtd)} = n^{O(\sqrtd)}$.

\mypara{The algorithm.}
We can now define a table entry $A[i,X]$ for each separator~$s_i$ and viable crossing pattern~$X$ at $s_i$ as follows.
\begin{quotation}
$A[i,X]$ :=  the minimum length of a rectilinear forest adhering to $X$ at $s_i$.
\end{quotation}
Note that the length of an MRST equals $A[n,\{\emptyset\}]$.
Next we describe a recursive formula to compute the table entries.
As a  base case, we will use $A[0, X] = 0$ for $X = \{\emptyset\}$, and $\infty$ for all other $X$.

Let $s_j$ and $s_i$ be consecutive separators, with $j < i$.
Note that since the point set is sparse, at most $k$ points share an $x$-coordinate.
Therefore, $j \geq i-k$.
Let $F(X,s_i)$ be a minimum-length rectilinear forest adhering to $X$ at $s_i$, and let $X'$ be its (unknown) crossing pattern at $s_j$.
Then the value of $A[i,X]$ equals the value of $A[j,X']$ plus the total length of the edges of $F(X,s_i)$ between $s_j$ and $s_i$.
The total length of $F(X,s_i)$ between these two separators only depends on $X'$ and $X$.
Since this subproblem contains $O(\sqrtd)$ points with three different $x$-coordinates, its Hanan grid contains only $O(\sqrtd)$ edges.
Therefore, its value can be computed in $2^{O(\sqrtd)}$ time by simply checking every possible subset of edges.
Let $L(X',X)$ denote the total length of the solution to this subproblem.
If no solution exists, we define it to be $\infty$.
Then we get
\[A[i,X] = \min_{\text{viable }X'} A[j,X'] + L(X',X),\]
where $s_j$ is the separator immediately preceding~$s_i$, and the sum is over all crossing patterns $X'$ that are viable at~$s_j$.

\mypara{The running time.}
To analyse the running time, we first determine the number of table entries.
There are $O(n)$ separators, and we have already seen for every separator $s_i$ there are $n^{O(\sqrtd)}$ possible viable crossing patterns.
Hence, the total number of table entries is $O(n) \cdot n^{O(\sqrtd)} = n^{O(\sqrtd)}$.
Next, we calculate the time needed per table entry.
For each of the $n^{O(\sqrtd)}$ possible viable crossing patterns $X'$ we compute $L(X',X)$ in $2^{O(\sqrtd)}$ time.
This brings the total time needed per table entry to $n^{O(\sqrtd)}$.

Since we have $n^{O(\sqrtd)}$ table entries, each needing $n^{O(\sqrtd)}$ time, we conclude:
\begin{theorem}
Let $P$ be a sparse point set of size $n$ inside a $\delta$-strip.
Then we can compute an MRST on $P$ in $n^{O(\sqrtd)}$ time.
\end{theorem}

Remark: The $n^{O(\sqrt{\delta})}$ running time of our algorithm is caused by the fact that we have bounded the number of viable crossing patterns at a given separator by $n^{O(\sqrt{\delta})}$.
One may wonder if the number of viable crossing patterns can really be that high.
Unfortunately the answer is yes: in the appendix we give an example of a point set where the number of viable crossing patterns is $f(\delta) \cdot n^{\Theta(\sqrt{\delta})}$, for some function~$f$.

\begin{proposition}\label{prop:sparse:xpatternlowerbound}
Let $n$ be large enough.
Then there exist a function $f$, a sparse point set $P_{\text{left}}$ of $n/2$ points and a set $\mathcal{P}$ of $f(\delta) \cdot n^{\Theta(\sqrtd)}$ sparse point sets $P_i$ of $\Theta(\sqrtd)$ points which lie fully to the right of $P_{\text{left}}$ such that for every $i$, all MRSTs on $P_{\text{left}} \cup P_i$ have the same crossing pattern at $s_n$, but this crossing pattern is different for all $i$.
\end{proposition}
For the proof, see Appendix~\ref{app:proofs}.

\section{Random point sets inside a narrow rectangle}\label{sec:random}
In this section we give an algorithm with $\min \{n^{O(\sqrtd)}, 2^{O(\delta \sqrtd)} n\}$ 
expected running time for points generated randomly inside a rectangle of height~$\delta$ and expected width~$n$.
Specifically, we assume the points in $P$ are generated as follows.
First, we generate $n$ independent exponentially distributed variables 
$\Delta_0,...,\Delta_{n-1} \sim \Exp (1)$. Using these, we compute the $x$-coordinates of our points
by setting $x_i := \sum_{j=0}^{i-1} \Delta_i$ for $1 \leq i \leq n$. 
Next, we generate the $y$-coordinates of the points by picking each $y_i$ uniformly 
and independently from the interval~$[0,\delta]$. 
Thus the points from $P$ lie inside the rectangle $[0,x_n]\times[0,\delta]$.
Since the spacings $\Delta_i$ are chosen from an exponential distribution of rate~1,
we have $\EE[x_n]=\EE[\sum_{i=0}^{n-1}\Delta_i]=n$. (More precisely, $\sum_{i=0}^{n-1}\Delta_i$
converges to a normal distribution with mean~$n$ and variance~$\sqrt{n}$.)

Recall that the algorithm for sparse point sets from the previous section, which had running time $n^{O(\sqrt{\delta})}$, was based on the fact that any separator $s_i$ of a sparse point set is crossed only $O(\sqrt{\delta})$ times.
Thus for each separator there are $n^{O(\sqrt{\delta})}$ different crossing patterns.
Our main goal is now to change this algorithm into an algorithm for random point sets that is fixed-parameter tractable with parameter~$\delta$.
We face two difficulties.
First, unlike in the case of sparse point sets, we cannot guarantee that all separators are crossed only $O(\sqrt{\delta})$ times.
Second, even if a separator is crossed $O(\sqrt{\delta})$ times, the number of candidate crossing patterns can still be $n^{\Theta(\sqrt{\delta})}$, which is too much for an FPT algorithm.
We overcome these difficulties as follows.

To deal with the first issue we will define a certain configuration of points and a corresponding separator---we will call such separator a \emph{soft wall}---such that the separator is crossed only $O(\sqrt{\delta})$ times.
Our new dynamic programming algorithm will have table entries for every soft wall instead of for every separator.
We will prove that we expect to find sufficiently many soft walls, so that the expected number of points in between two consecutive soft walls only depends on~$\delta$ (and not on~$n$).
This still leaves the second problem, because where a soft wall is crossed by an MRST may depend on points from $P$ that are beyond the previous or next soft wall.
Thus the number of crossing patterns can still be $n^{\Theta(\sqrt{\delta})}$.
We therefore also devise a second type of wall, the \emph{hard wall}.
This is a vertical line~$\ell$ through an input point~$p_i$ that will not be crossed at all by an edge of
an MRST.
The MRST will consist of two independent parts:
an MRST for the points to the left of $\ell$ plus $p_i$ itself, and an MRST for the points to the right of $\ell$ plus $p_i$ itself.
More generally, if we have a collection of hard walls then the subproblems between any two consecutive hard walls are completely independent.
Hard walls will occur much less frequently than soft walls,
but still the expected number of points in between two consecutive hard walls will be shown to depend only on~$\delta$.
Hence, the number of crossing patterns we need to consider for the soft walls in between the two hard walls only depends on $\delta$, giving us an FPT algorithm.

See Algorithm~\ref{alg:random} for pseudocode for the global algorithm.
Recall that $P[i,j] \mydef \{p_i,...,p_j\}$.
The constant 100 mentioned is not special; it is merely an arbitrary large enough constant.

\begin{algorithm} 
\caption{\textsc{ComputeMRST}$(P)$} \label{alg:random}
\begin{algorithmic}[1]
\State Compute a collection $\hwalls = \{\ell_0,\ldots,\ell_m\}$ of hard walls, as described below.
       The walls in $\hwalls$ are numbered from left to right, with $\ell_0$ and $\ell_m$ 
       being `hard walls' consisting of the leftmost and rightmost points of $P$, respectively.
\For{$i \gets 0$ to $m-1$} 
    \State Let $p_j$ and $p_{j'}$ be the middle points of the hard walls $\ell_i$ and $\ell_{i+1}$, respectively.
    \If{$\delta < 100$}
        \State Compute an MRST~$T_i$ for $P[j,j']$ using the $2^{O(\sqrt{n} \log n)}$ algorithm by Fomin \etal
    \Else
        \State \begin{minipage}[t]{123mm} Compute a collection $\swalls=\{t_1,\ldots,t_z \}$ of soft walls for $P[j,j']$, as described below.  \vspace{1.75mm} \end{minipage}
        \State \begin{minipage}[t]{123mm}
                Compute an MRST~$T_i$ for $P[j,j']$ using the dynamic-programming algorithm described in
                Section~\ref{sec:sparse}, but using the collection $\swalls$ as separators (instead of
                using all separators between consecutive points), as described below.
            \end{minipage}
    \EndIf
\EndFor
\State \Return $T_0\cup\cdots \cup T_{m-1}$.
\end{algorithmic}
\end{algorithm}

\mypara{Computing hard walls.}
Let $P[i,i+4]$ be a subset of points from $P$, and let $\ell$ be the vertical 
line through~$p_{i+2}$. We call $\ell$ a \emph{hard wall} if $P[i,i+4]$ has the following 
properties:
\begin{itemize}
    \item $\Delta_j > \delta$ for all $i\leq j\leq i+3$
    \item $y_{i+1} < y_{i+2} < y_{i+3}$
\end{itemize}
\begin{figure}
\begin{center}
\includegraphics{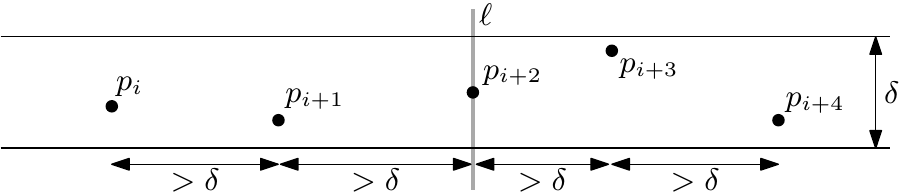}
\caption{Example of a hard wall}
\label{fig:random:hardwall}
\end{center}
\end{figure}

\noindent See Figure~\ref{fig:random:hardwall} for an example of a hard wall.
A hard wall indeed splits the problem into independent subproblems, as shown in the lemma below.
\begin{lemma}\label{obs:random:hardwall}
Let $\ell$ be a hard wall, defined by the subset~$P[i,i+4]$.
Let $T_1$ be an MRST on $P[1,i+2]$ and let $T_2$ be an MRST on $P[i+2,n]$.
Then $\|T\| = \|T_1\| + \|T_2\|$ and so $T_1 \cup T_2$ is an MRST on $P$.
\end{lemma}
\begin{proof}
Let $T$ be a Hwang tree on~$P$.
By Observation~\ref{obs:gen:paralleledges} we know that an MRST on $P$
is monotonic at $s_i,...,s_{i+3}$. The monotonicity at $s_{i+1}$ and $s_{i+2}$
implies that $\degr_T(p_{i+2})\leq 2$. 
If $\degr_T(p_{i+2})=2$ then splitting $T$ at $p_{i+2}$ results
in subtrees on $P[1,i+2]$ and $P[i+2,n]$---this follows from the
monotonicity at $s_{i+1}$ and $s_{i+2}$---and so we are done. 
Now assume for a contradiction that $\degr_T(p_{i+2})=1$. Then the incident edge
if $p_{i+2}$ must be vertical. Assume without loss of generality that $p_{i+2}$
is the top endpoint of this edge. But then the (single) edge of $T$ crossing $s_{i+2}$ must reach the vertical
line through $p_{i+3}$ at a point $q$ that lies somewhere below~$p_{i+3}$. The
monotonicity at $s_{i+3}$ then implies that $q$ must be connected
to $p_{i+3}$ by a vertical segment, thus creating a U-shape and contradicting that $T$ is a Hwang tree.
See Figure~\ref{fig:random:hardwall2} for an example.
\begin{figure}
\begin{center}
\includegraphics{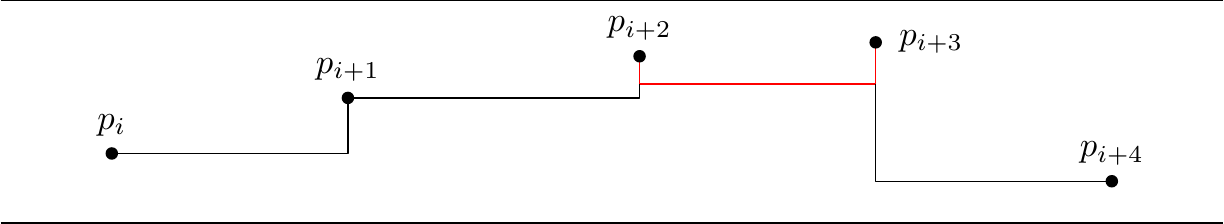}
\caption{Illustration for the proof of Lemma~\ref{obs:random:hardwall} showing an MRST $T$ where $\degr_T(p_{i+2})=1$.
In red, the $U$-shape showing that $T$ is not a Hwang tree.}
\label{fig:random:hardwall2}
\end{center}
\end{figure}
\end{proof}
The next lemma gives a bound on the probability that $P[i,i+4]$ is a hard wall.
\begin{lemma}\label{le:hard-wall}
$\PP\Big[\; \mbox{$P[i,i+4]$ defines a hard wall} \;\Big] \ = \ e^{-4 \delta}/6$ \ \ for all $1\leq i\leq n-4$.
\end{lemma}
\begin{proof}
Recall that the spacings $\Delta_j$ are drawn independently from an exponential distribution with rate~1.
Hence, $\PP[\Delta_j]>\delta]=e^{-\delta}$ for all $j$. Since the spacings are independent,
the probability that all four spacings between the points in $P[i,i+4]$ are greater than~$\delta$
is~$e^{-4\delta}$. Finally, $\PP[y_{i+1}<y_{i+2}<y_{i+3}]=1/6$, since all $y$-coordinates
are chosen uniformly at random from~$[0,\delta]$ and so all six 
orderings of $y_{i+1}<y_{y+2}<y_{i+3}$ are equally likely.
\end{proof}
The set $\hwalls$ of hard walls is now computed in the following straightforward manner:
we check for all $i:=5j+1$ with $j\in\{0,\ldots,\floor{n/5}-1\}$ whether $P[i,i+4]$ 
defines a hard wall; if so, we add the corresponding hard wall to~$\hwalls$.
Note that this takes only $O(n)$ time in total, as each of the $O(n)$ candidate hard walls can be checked in $O(1)$ time.

\begin{figure}[b]
\begin{center}
\includegraphics{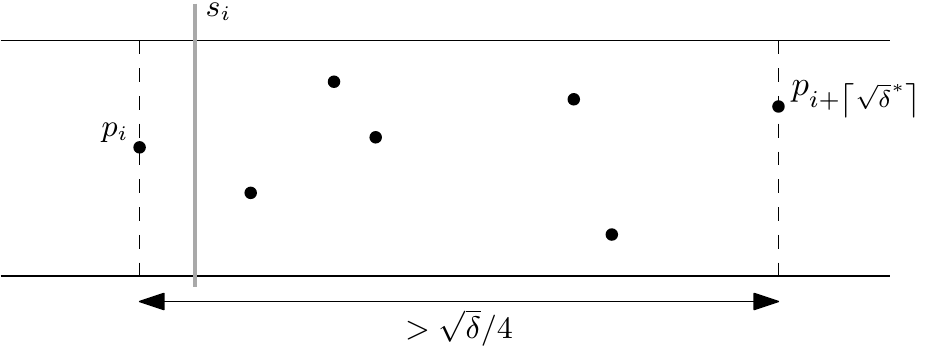}
\caption{Example of a soft wall}
\label{fig:random:softwall}
\end{center}
\end{figure}

\mypara{Computing soft walls.}
Let $P[i,i+\cisqrtd]$ be a subset of $\cisqrtd+1$ points from~$P$ such 
that~$x_{i+\cisqrtd} - x_i > \cisqrtd/4$.
Then we call the separator~$s_i$---recall
that $s_i$ is the separator between $p_i$ and $p_{i+1}$---a \emph{soft wall}.
See Figure~\ref{fig:random:softwall} for an example.
\begin{lemma}\label{le:soft-wall}
Let $\delta \geq 100$.
Let $s_i$ be a soft wall, defined by $P[i,i+\cisqrtd]$.
Then $s_i$ is crossed $O(\sqrt{\delta})$ times by an MRST. 
Furthermore, even under the assumption that $\Delta_j < \delta$ for all $1 \leq j \leq n-1$,
we have
\[
\PP\sbr{\; \mbox{$P\sbr{i,i+\csqrtd}$ defines a soft wall} \;} \ \geq \ 1 - 2^{3 - \csqrtd/2}
\mbox{ for all $1\leq i\leq n-\csqrtd$.}
\]
\end{lemma}
\begin{proof}
The fact that $s_i$ is crossed at most $O(1+\cisqrtd)=O(\cisqrtd)$ times follows immediately
from Lemma~\ref{lem:gen:sqrtdeltacrossings}.
To be precise, $s_i$ is crossed at most $18(2+ \sqrtd)$ times.
It remains to derive a lower bound on  
the probability that $P[i,i+\cisqrtd]$ is a soft wall, given that $\Delta_j < \delta$ for all $1 \leq j \leq n-1$.
We have
\begin{align*}
& \PP \left[ x_{i+\csqrtd} - x_i > \csqrtd/4 \right] \\
& = 1 - \PP \sbr{ \sum_{j=i}^{i+\csqrtd-1} \Delta_j \leq \csqrtd/4 }\\
& \geq 1 - \min_{t>0} e^{t \csqrtd/4} \br{ \EE[e^{-t\Delta_1}]}^{\csqrtd-1}
& \text{by the Chernoff bound}\\
& \geq 1 - \min_{t>0} e^{t \csqrtd /4} \br{ \int_{x=0}^{\delta} e^{-tx}e^{-x} \frac{1}{1-e^{-\delta}} }^{\csqrtd-1}
& \Delta_1 \sim \Exp(1) \text{ and } \Delta_1 \leq \delta
\end{align*}
Now, we can simply pick $t= 3$ to obtain
\begin{align*}
& 1 - \min_{t>0} e^{t \csqrtd /4} \br{ \int_{x=0}^{\delta} e^{-tx}e^{-x} \frac{1}{1-e^{-\delta}} }^{\csqrtd-1} \\
& \geq 1 - e^{3 \csqrtd /4} \br{ \int_{x=0}^{\delta} e^{-3x}e^{-x} \frac{1}{1-e^{-\delta}} }^{\csqrtd-1} \\
& = 1 - e^{3 \csqrtd/4} \br{\frac{1}{4}(1+e^{-\delta})(1+e^{-2 \delta})}^{\csqrtd-1}\\
& > 1 - 2e^{3\csqrtd/4} 4^{1-\csqrtd}\\
& > 1 - 2^{3 - \csqrtd/2} \qedhere
\end{align*}
\end{proof}
Recall that in Algorithm~\textsc{ComputeMRST} we need to compute soft walls for every subset
$P[j,j']$ between two consecutive hard walls (including the points on those two hard walls).
To this end we check whether $P[p_r,...,p_{r+\cisqrtd}]$ forms a soft wall for all
$r:= j+i \cisqrtd $ with $0\leq i\leq j'-\cisqrtd$.

\begin{figure}
\begin{center}
\includegraphics{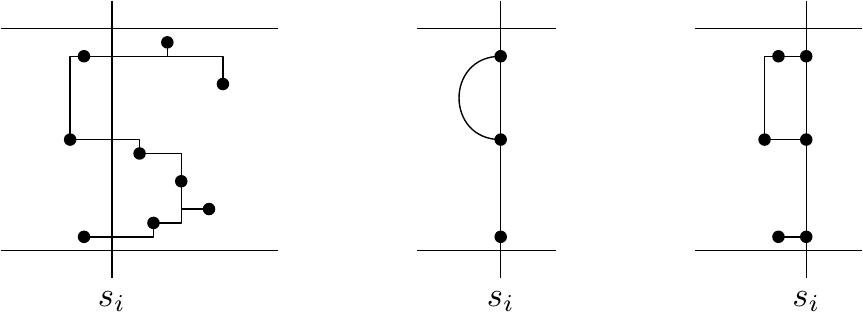}
\caption{An example of an element of a table entry $A[i]$.
On the left, an MRST $T$ represented in $A[i]$.
$T$ is represented in $A[i]$ by a pair $(X_i,l)$.
In the middle, we have $X_i$.
On the right, the edges contributing to the length $l$.
Note that since $T$ is an MRST, these edges indeed form a minimal length rectilinear forest adhering to $X_i$ at $s_i$.}
\label{fig:random:randomalgoA}
\end{center}
\end{figure}

\mypara{The dynamic-programming algorithm between two hard walls.}
Recall that $X(s_i)$ is the set of points where the Hanan grid crosses $s_i$.
Let $\mathcal{X}_i$ denote the family of subsets of $X(s_i)$ of size at most $18(2+ \sqrtd)$.
We can now define a table entry $A[i]$ for each \emph{soft wall}~$s_i$ as follows.
\begin{quotation}
$A[i]$ :=  a representative set of pairs $(X, l)$ where $l$ is the minimum length of a 
           \\ \hspace*{16mm} rectilinear forest adhering to $X \in \mathcal{X}_i$ at $s_i$.
\end{quotation}
Here, `representative' means that for every soft wall $s_i$ there exists an MRST $T$ 
and $(X,l) \in A[i]$ such that $T$ adheres to $X$ at $s_i$.
We will call this $T$ an MRST \emph{represented in} $A[i]$.
See Figure~\ref{fig:random:randomalgoA} for an example.
Note that $A[n]$ contains one element, of which the length~$l$ equals the length of an MRST on $P$.
Next we describe a recursive formula to compute the table entries.
As base case, we have $A[0] = \{\{\emptyset\},0\}$.
We first give pseudocode for this part of the algorithm.

\begin{algorithm}[h] 
\caption{\textsc{ComputeA}$(i)$} \label{alg:randominside}
\begin{algorithmic}[1]
\State Let $s_j$ be the rightmost soft wall to the left of $s_i$.
\For{$(X_j, l) \in A[j]$}
    \For{all viable mirrored crossing patterns $X_i'$ at $s_i$}
        \State Compute an MRST $T'$ for the subproblem given by $X_j$ and $X_i$.
        \State Add $(X_i, l + \|T'\|)$ to $A[i]$, where $X_i$ is the crossing pattern of $T'$ at $s_i$.
    \EndFor
\EndFor
\State Remove unviable pairs from $A[i]$.
\end{algorithmic}
\end{algorithm}

Let $s_i$ be a soft wall, and let $s_j$ be the rightmost soft wall to the left of $s_i$.
We define a mirrored crossing pattern to be a crossing pattern where the partition denotes on how the rectilinear Steiner tree is connected on the \emph{right} hand side.
Let $(X_j,l)$ be a pair in $A[j]$.
Let $T$ be an MRST adhering to $X_j$ at $s_j$, and adhering to some (unknown) mirrored crossing pattern $X_i'$ at $s_i$.
Then there is a pair $(X_i,l')$ in $A[i]$, where $l'$ equals $l$ plus the total length of the edges of $T$ between $s_j$ and $s_i$.
The total length of $T$ between these two separators only depends on $X_j$ and $X_i'$.
Let $L(X_j,X_i')$ denote the total length of the solution to this subproblem.
Now, to compute the value of $L(X_j, X_i')$, we use the $2^{O(\sqrt{n} \log n)}$ algorithm by Fomin~\etal~\cite{rectiSteiner16}.
Since this algorithm only computes Steiner trees (not forests adhering to some crossing
pattern), we need to adapt our subproblem.
To ensure the crossing pattern $X_j$, we mimic the edges on the left of $X_j$.
For every part of $X_j$, we add a path of `virtual' edges of length $0$, connecting the points in that part.
These are automatically added to the so-called shortest path RST found by the first part of the algorithm by Fomin~\etal\
Since the number of virtual edges added is constant in $n$, it does not affect its running time.
We ensure the crossing pattern $X_i'$ analogously.
Given the output $T'$ of the algorithm, we remove its virtual edges, and analyse its (non-mirrored) crossing pattern $X_i$ at $s_i$.
We then add the pair $(X_i, l+\|T'\|)$ to $A[i]$.
After doing so for all pairs in $A[j]$ and viable mirrored crossing patterns $X_i'$, we may be able to remove some elements from $A[i]$.
First, we remove any duplicates.
Then, if two pairs have the same crossing pattern $X_i$, we need only the one with the smallest $l$.

\begin{figure}
\begin{center}
\includegraphics{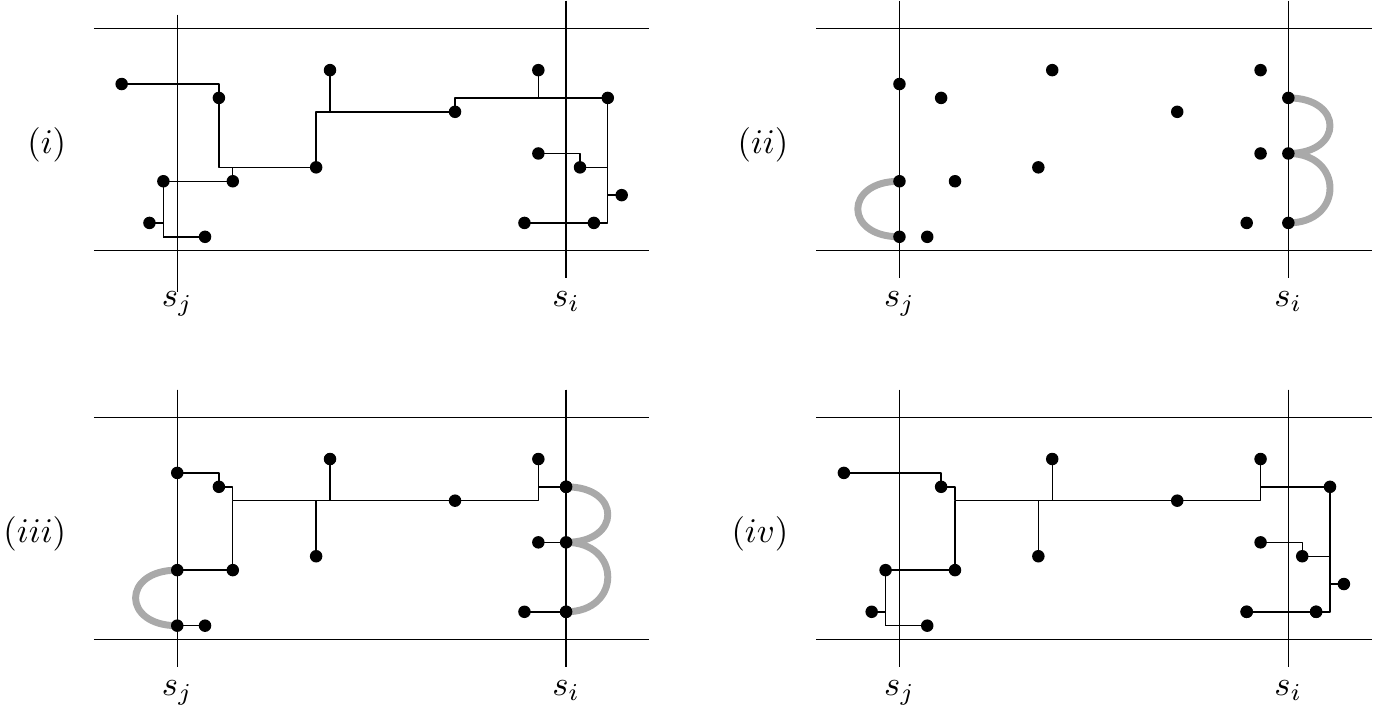}
\caption{Illustration for the correctness proof of Algorithm~\ref{alg:randominside}. At (i), we have an MRST $T$ which adheres to $X_j$ at $s_j$, and to $X_i'$ at $s_i$. At (ii), we have the corresponding subproblem. The thick grey edges denote the virtual edges of length 0. At (iii), a solution $T'$ to the subproblem. Note that the total length of $T'$ equals the total length of $T$ between $s_j$ and $s_i$. At (iv), the new MRST $T''$ represented in $A[i]$, obtained by combining $T$ and $T'$.}
\label{fig:random:randomalgo}
\end{center}
\end{figure}

We will now prove that $A[i]$ is indeed a representative set by induction on $i$.
Clearly, $A[0]$ is a representative set.
Now, suppose $A[j]$ is a representative set.
We will now show that after performing the above, $A[i]$ is a representative set.
See Figure~\ref{fig:random:randomalgo} for an example.
Since $A[j]$ is a representative set, there exists a pair $(X_j, l) \in A[j]$ and an MRST $T$ such that $T$ adheres to $X_j$ at $s_j$.
Now, $T$ adheres to some mirrored crossing pattern $X_i'$ at $s_i$.
Therefore, we will find an MRST $T'$ on the subproblem defined by $X_j$ and $X_i'$, and add a pair $(X_i,l')$ to $A[i]$.
Let $T''$ be the MRST on $P$ obtained by exchanging the part of $T$ between $s_j$ and $s_i$ for $T'$.
Note that we can do that, since $T$ and $T'$ adhere to the same crossing patterns $X_j$ and $X_i'$.
Now, $T''$ is represented in $A[i]$ by $(X_i, l')$.

\mypara{Analysis of the running time.}
We now analyze the expected running time of Algorithm~\textsc{ComputeMRST}.
To do so, we will bound certain distributions by other distributions.
To be precise, we bound the expected running time of any algorithm on a point set with a 
random number of points following a certain distribution by the expected running time of 
the algorithm on a point set with a differently distributed random number of points.
\begin{observation}[\cite{DBLP:conf/compgeom/AlkemaBK20}] \label{obs:random:probbound}
Let $Y_1,Y_2$ be two discrete nonnegative random variables, such that for all $k \geq 0$, the equation $\PP[Y_1 \leq k] \geq \PP[Y_2 \leq k]$ holds.
Let $f(k)$ be an increasing nonnegative function such that $\EE[f(Y_2)] < \infty$.
Then
\[\EE[f(Y_1)] = \sum_{k=0}^\infty f(k) \PP[Y_1 = k] \leq \sum_{k=0}^\infty f(k) \PP[Y_2 = k] = \EE[f(Y_2)].\]
\end{observation}
We write $Y_1 \preceq Y_2$  to denote that for all $k \geq 0$, the equation $\PP[Y_1 \leq k] \geq \PP[Y_2 \leq k]$ holds.

Let us take a look at the sizes of the subproblems defined by the hard walls.
Suppose we are computing $\mathcal{W}_{hard}$ and have just found a hard wall $\ell_i$.
Let the random variable $X_1$ denote the number of points in the subproblem $P[j,j']$ between the two hard walls $\ell_i$ and the unknown $\ell_{i+1}$.
Note that $X_1$ is at most $m \mydef n-j+1$, and that $X_1$ only depends on $m$.
Therefore, we will write $X_{1,m}$.
Now, $X_{1,m}$ is almost geometrically distributed.
There are two differences: we only check whether $P[i,i+4]$ defines a hard wall for $i$ of the form $5j+1$, and $X_{1,m}$ is at most $m$.
Since the probability that $P[i,i+4]$ defines a hard wall is $e^{-4\delta}/6$, we have $X_{1,m} \sim \min\{m, 1 + 5 \cdot \text{Geom}(e^{-4\delta}/6)\}$.
Here, the probability mass function of $\text{Geom}(p)$ is $(1-p)^{k-1} p$.
Let $X_2$ be the same distribution, but where we ignore the maximum number of points, $X_2 \sim 1 + 5 \cdot \text{Geom}(e^{-4\delta}/6)$.
Then, $X_{1,m} \preceq X_2$ for all $m$.

We are now ready to calculate the expected running time of \textsc{ComputeMRST} if $\delta < 100$.
We have already seen that we can find $\hwalls$ in $O(n)$ time.
Since $X_{1,m} \preceq X_2$ for all $m$, the expected time needed per subproblem is bounded by the expected time needed to run the $2^{O(\sqrt{n}\log n)}$ algorithm by Fomin~\etal~on a point set with $X_2$ points. We get:
\[\sum_{k=1}^\infty (1-e^{-4\delta}/6)^{k-1} \cdot \br{e^{-4\delta}/6} \cdot 2^{O(\sqrt{5k+1} \log (5k+1))} < \sum_{k=1}^\infty 2^{-\Theta(k)} \cdot 2^{O(\sqrt{5k} \log (5k))} = O(1)\]
Since there are $O(n)$ subproblems, this finishes the case $\delta < 100$.

We can use the same trick for the distribution of the number of points between soft walls.
Here, we let the random variable $Y_{1,m}$ denote the number of points between two consecutive soft walls, given that we have found no hard walls between the hard walls defining our subproblem and where $m$ is once more the maximum number of points.
We can bound $Y_{1,m}$ in three steps.
First, note that the condition that no $\Delta_j$ is larger than $\delta$ is stronger than the condition that there are no hard walls between the hard walls defining our subproblem.
Let $Y_{2,m}$ denote the number of points between the soft walls, given that no $\Delta_j$ is larger than $\delta$.
Then $Y_{1,m} \preceq Y_{2,m}$.
Next, recall that if $\delta \geq 100$, by Lemma~\ref{le:soft-wall} the probability that $s_i$ is a soft wall is at least $1 - 2^{3 - \cisqrtd/2}$, even if all $\Delta_j<\delta$.
Define $Y_{3,m} \sim \min\{m, \cisqrtd \cdot \text{Geom}(1 - 2^{3 - \cisqrtd/2})\}$.
Then $Y_{2,m} \preceq Y_{3,m}$.
Finally, analogously to the hard walls, we can remove the maximum number of points.
Define $Y_4 \sim \cisqrtd \cdot \text{Geom}(1 - 2^{3 - \cisqrtd/2})$.
We conclude that $Y_{1,m} \preceq Y_4$ for all $m$.

Let $c, \labda > 0$ be such that the algorithm by Fomin~\etal~runs in under $c \cdot 2^{\labda \sqrt{n} \log n}$ time.
For the case $\delta \geq 100$, the total expected time needed per subsubproblem is then bounded by
\begin{align*}
    &\sum_{i=1}^\infty c \cdot 2^{\labda \sqrt{i \csqrtd +1}\log \br{i \csqrtd+1}} \cdot \br{2^{3-\csqrtd/2}}^{i-1} \cdot \br{1 - 2^{3-\csqrtd/2}} \\
    &< \; c \cdot 2^{\csqrtd/2-3} \sum_{i=1}^\infty 2^{\labda \sqrt{2i \csqrtd}\log \br{2i \csqrtd}} \cdot 2^{(3-\csqrtd/2)i} \cdot 1\\
    &< \; c \cdot 2^{\csqrtd} \sum_{i=1}^\infty 2^{4 \labda (2 i\csqrtd)^{3/4}} \cdot 2^{-i \csqrtd/5} 
    & \text{since } \delta \geq 100\\
    &< \; c \cdot 2^{\csqrtd} \sum_{i=1}^\infty 2^{\csqrtd  (8 \labda i^{3/4} - i / 5)}
\end{align*}
Now, for a sufficiently large~$M$, we have $8 \labda i^{3/4} - i / 5 < -i/10$ for all $i\geq M$.
We get:
\begin{align*}
    & c \cdot 2^{\csqrtd} \sum_{i=1}^\infty 2^{\csqrtd  (8 \labda i^{3/4} - i / 5)} \\
    &= \; c \cdot 2^{\csqrtd} \br{ \sum_{i=1}^{M} 2^{\csqrtd  (8 \labda i^{3/4} - i / 5)} + \sum_{i=M+1}^\infty 2^{\csqrtd  (8 \labda i^{3/4} - i / 5)}} \\
    &< \; c \cdot 2^{\csqrtd} \br{ M \cdot 2^{\csqrtd \max_{i \geq 1} (8 \labda i^{3/4} - i / 5)} + \sum_{i=M+1}^\infty 2^{- i / 10}} 
    = \; 2^{O\br{\csqrtd}}.
\end{align*}
Let $m$ be the number of points in the corresponding subproblem defined by two hard walls.
Note that the above bound is independent of $m$.
Analogously to the original sparse point-set algorithm, there are $m^{O(\csqrtd)}$ possible crossing patterns per separator.
In total, this algorithm therefore takes $m \, \cdot m^{O(\csqrtd)} \cdot 2^{O\br{\csqrtd}} = m^{O(\sqrtd)}$ expected time.

Now, all that remains is calculating the expected running time of our main random point set algorithm in this case.
Clearly, it runs in at most expected $n^{O(\sqrtd)}$ time, since splitting up the problem using the hard walls can only speed up the algorithm.

Let $\labda, \mu \geq 1$ be such that the $m^{O(\sqrtd)}$ expected running time algorithm runs in at most $m^{\labda \sqrtd}$ expected time, and that the probability of a hard wall is $2^{- \mu \delta}$.
Let $Y_1$ be the distribution of the number of points of a subproblem.
Recall that $Y_1 \preceq 1 + 5 \cdot \text{Geom}(e^{-4\delta}/6)$.
Then the total expected time needed per subproblem is bounded by
\[
\EE \sbr{ Y_1^{\labda \sqrtd} }
\leq \sum_{k=1}^\infty (1-2^{-\mu \delta})^{k-1} \cdot 2^{-\mu \delta} \cdot \br{5k+1}^{\labda \sqrtd} < O(1) + 2^{-\mu \delta} \sum_{k=2}^\infty 2^{4\labda \sqrtd \log k -2^{-\mu \delta} k} \]
We split the sum into two parts, with $M=64 \labda^2 \delta 2^{2 \mu \delta}$. We get
\begin{align*}
&2^{-\mu \delta} \sum_{k=2}^\infty 2^{4\labda \sqrtd \log k -2^{-\mu \delta} k}\\
&< 2^{-\mu \delta} \br{ \sum_{k=2}^{M-1} 2^{4\labda \sqrtd \log k -2^{-\mu \delta} k} +  \sum_{k=M}^\infty 2^{-2^{-\mu \delta} k/2} }\\
&< 2^{-\mu \delta} M 2^{\max_{k} \br{\labda \sqrtd \log k -2^{-\mu \delta} k}} + \frac{2^{-\mu \delta}}{1-2^{-2^{-\mu \delta-1}}} \\
&= 2^{-\mu \delta} M 2^{\labda \sqrtd \log (\labda \sqrtd 2^{\mu \delta} \ln 2) -2^{-\mu \delta} (\labda \sqrtd 2^{\mu \delta} \ln 2)} + O(1) = 2^{O(\delta \sqrtd)},
\end{align*}
since $\frac{2^{-\mu \delta}}{1-2^{-2^{-\mu \delta-1}}}$ converges to $2 / \log(2)$.
This brings the total expected running time to $O(n) \cdot (O(1) + 2^{O(\delta \sqrtd)}) = 2^{O(\delta \sqrtd)} n$.

All in all, our main random point set algorithm run in $O(n)$ expected time if $\delta < 100$, and in $\min\{n^{O(\sqrtd)},2^{O(\delta \sqrtd)} n\}$ expected time if $\delta \geq 100$.
We conclude:
\begin{theorem}
Let $P$ be a set of $n$ points generated randomly inside a rectangle of height~$\delta$
and expected width~$n$, generated according to the procedure described earlier.
Then an MRST on $P$ can be found in $\min\{n^{O(\sqrtd)},2^{O(\delta \sqrtd)} n\}$ expected time.
\end{theorem}

\section{Concluding remarks}
Our paper contains two main results on \mrst.
First, we proved that for sparse point sets in a strip of width~$\delta$, an MRST can be found in $n^{O(\sqrtd)}$ time.
Second, we gave a $\min\{n^{O(\sqrtd)},2^{O(\delta \sqrtd)} n\}$ expected running time algorithm for random point sets.
For $\delta=\Theta(n)$ the running time equals the $2^{O(\sqrt{n} \log n)}$ of the algorithm for arbitrary point sets in the plane~\cite{rectiSteiner16}.
A challenging open problem is to see if an algorithm with running time~$2^{O(\sqrt{\delta}\log\delta)}\mbox{poly}(n)$ is possible.
Another direction for future research is to study the problem in higher dimensions.
We believe that our algorithmic results may carry over to $\Reals^d$ to points that are almost collinear, that is, that lie in a narrow cylinder.
Generalizing the results to, say, points lying in a narrow slab will most likely be more challenging.

More generally, we believe that it is interesting to study the parameterized complexity of geometric problems using a ``geometric parameter''.
For problems involving planar point sets, the strip width $\delta$ is a natural parameter, which is interesting because it explores the boundary between the 1-dimensional and 2-dimensional version of the problem.
We have studied this for TSP in a previous paper~\cite{DBLP:conf/compgeom/AlkemaBK20} and for \textsc{Minimum Rectilinear Steiner Tree} in the current paper, but many other problems can be studied from this perspective as well.

\section*{Acknowledgements}
We thank Remco van der Hofstad for discussions about the probabilistic analysis.

\bibliography{bibfile}

\appendix

\newpage

\section{Proof of Proposition~\ref{prop:sparse:xpatternlowerbound}}\label{app:proofs}
\textit{Let $n$ be large enough.
Then there exist a function $f$, a sparse point set $P_{\text{left}}$ of $n/2$ points and a set $\mathcal{P}$ of $f(\delta) \cdot n^{\Theta(\sqrtd)}$ sparse point sets $P_i$ of $\Theta(\sqrtd)$ points which lie fully to the right of $P_{\text{left}}$ such that for every $i$, all MRSTs on $P_{\text{left}} \cup P_i$ have the same crossing pattern at $s_n$, but this crossing pattern is different for all $i$.}
\begin{proof}
We will show that this indeed holds for $f(\delta) \mydef \delta^{-\Theta(\sqrtd)}$.
We start by introducing a gadget, which we will call a \emph{hook}.
A hook $H$ is a set of $m$ points in a `$<$'-form, where the points (ordered from left to right) alternately are above the highest point and below the lowest point so far.
See Figure~\ref{fig:misc:hook} for an example.
Suppose we add a point $p$ which is to the right of $H$ and has a $y$-coordinate used by a point of $H$.
Then, any MRST on the set $H \cup \{p\}$ contains an edge from $p$ going left.
Therefore, we can use a hook of $m$ points to generate $m$ different crossing patterns.
Note that the difference in $y$-coordinates of the points can be arbitrarily small, so we will treat them as such.
\begin{figure}
\begin{center}
\includegraphics{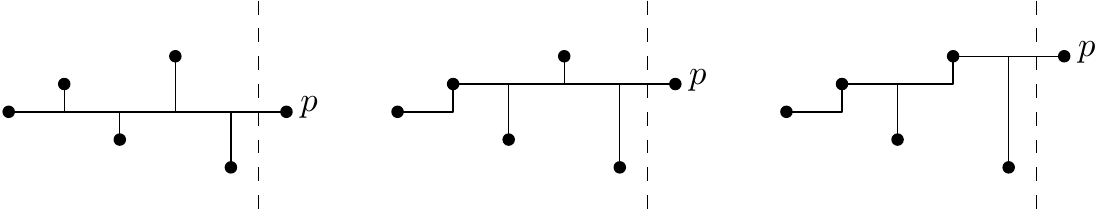}
\caption{An example of a hook and three of the five resulting crossing patterns ($y$-coordinates exaggerated for visibility).}
\label{fig:misc:hook}
\end{center}
\end{figure}

We will now place $\Theta(\sqrtd)$ hooks containing $\Theta(n / \sqrtd)$ points each below each other.
If these hooks `act' independently, we can choose $\Theta(n/\sqrtd)$ different $p$ for each of the $\Theta(\sqrtd)$ hooks, resulting in the required $\Theta(n / \sqrtd)^{\Theta(\sqrtd)} = n^{\Theta(\sqrtd)} f(\delta)$ crossing patterns.
To do so, we connect the hooks on the left hand side.
See Figure~\ref{fig:misc:hook2} for an example.
\begin{figure}
\begin{center}
\includegraphics{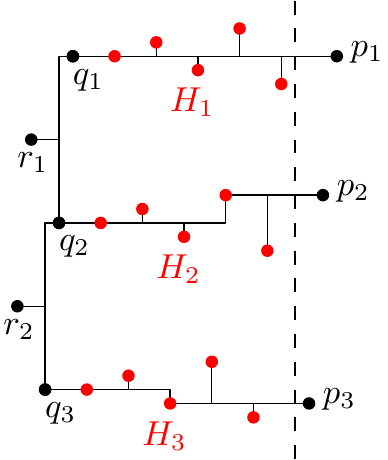}
\caption{An example of a point set $P \cup P_i$. $P$ to the left of the separator, the three hooks of $P$ in red, and $P_i = \{p_1,p_2,p_3\}$ to the right of the separator. Scaled for visibility; the vertical distance between two hooks is approximately $100$ times larger than the horizontal distance between two points of a hook.}
\label{fig:misc:hook2}
\end{center}
\end{figure}
To be precise, suppose we have $m = \ceili{\sqrtd / 10}$ hooks $H_1, ..., H_m$ from top to bottom.
To ensure the point set is sparse, all points will have distinct integer $x$-coordinates.
We distribute these equally such that for each hook, the horizontal distance between two consecutive points is $m$.
The offsets are such that $H_i$ is more to the right than $H_j$ if $i < j$.
Note that the vertical distance between two hooks is approximately $\delta / (m-1) \approx 10 \sqrtd \approx 100m$.
To the left of each hook $H_i$ we add an extra points $q_i$ on the same height of the leftmost point of $H_i$.
Finally, for each $i < m$, we add a point $r_i$ to the left of $H_i$ with a height halfway between those of the leftmost points of $H_i$ and $H_{i+1}$.
These points together form the point set $P$.

Now, all that remains is to show that the hooks indeed `act' independently.
This is easy to see; as the vertical distance between two hooks is a factor $100$ larger than the horizontal distance between two consecutive points in the same `row', the vertical distance between two consecutive hooks will be bridged exactly once.
Since there must be a horizontal segment between the leftmost point of every $H_i$ and the corresponding $q_i$, every $r_i$ is connected to $q_i$ and $q_{i+1}$.
Since the hooks are guaranteed to be connected on the left hand side, each of the $f(\delta) \cdot n^{\Theta(\sqrtd)}$ different combinations of points $p_1, ..., p_m$ guarantees a different crossing pattern, as required.
\end{proof}


\end{document}